\documentclass[a4paper,10pt]{article}
\usepackage{theorem,amssymb}
\usepackage{graphicx}
\def\eps{\varepsilon}
\def\bE{{\mathbb E}}
\def\Var{\mathrm{Var}\,}

\def\<{\langle}
\def\>{\rangle}

\begin{document}

\title{Estimation of nonclassical independent Gaussian processes by classical interferometry}

\author{L\'aszl\'o Ruppert$^{1,*}$, Radim Filip$^{1,+}$\\
\small $^1$Department of Optics, Palacky University, 17. listopadu 12, 771 46 Olomouc, Czech
Republic\\
\small\textit{$^*$ruppert@optics.upol.cz, $^+$filip@optics.upol.cz}}
\date{}


\maketitle

\begin{abstract}
We propose classical interferometry with low-intensity thermal radiation for the estimation of nonclassical independent Gaussian processes in material samples. We generally determine the mean square error of the phase-independent parameters of an unknown Gaussian process, considering a noisy source of radiation the phase of which is not locked to the pump of the process. We verify the sufficiency of passive optical elements in the interferometer, active optical elements do not improve the quality of the estimation. We also prove the robustness of the method against the noise and loss in both interferometric channels and the sample. The proposed method is suitable even for the case when a source of radiation sufficient for homodyne detection is not available.
\end{abstract}

\section{Introduction}

The development of quantum technology in various platforms of physics highly depends on the experimental ability to generate and detect nonclassical quantum processes for electromagnetic radiation. Nonclassical processes can change classical radiation to radiation incompatible with any mixture of classical waves \cite{Glauber}. 
To witness phase-independent nonclassical processes, very sensitive photon counting measurements can be used \cite{Grangier,hom,Jezek}. Such photon counters can observe single-photon production arising from weak quantum nonlinear processes \cite{anti1,anti2,anti3,anti4,hom1,hom2,hom3,hom4}.
The next step of investigation is to witness phase-dependent aspects of these nonclassical processes. For the lowest order of nonlinear effects, which dominate in weakly nonlinear processes enhanced by a strong pump, the squeezing of continuous amplitude fluctuations of radiation for a given phase is the dominant nonclassical process \cite{sqexp1,sqexp2}. It is an important process because the squeezing of quantum fluctuations of radiation has many direct applications in modern quantum technology.

To estimate phase-sensitive squeezing, we need to probe the non-classical processes by a suitable wavelength of radiation capable of classical interference. Nowadays, we can use narrow-band sources of radiation at various wavelengths to probe many new nonlinear processes in samples of different materials or soft matter. A standard narrow band source for spectroscopy typically exhibits thermal statistics with limited energy per time duration of the measurement. Many of such sources can be sufficient to build classical interferometry, for example, to estimate the relative classical phase introduced by the sample \cite{BW}. This is because a limited intensity of thermal light is sufficient for classical interferometry, which requires only coherence in first order irrespective of the noise of the light \cite{MW}.

On the other hand, the estimation of squeezing in material samples has been limited to balanced homodyne detection (BHD) \cite{bhom1,bhom2} and unbalanced homodyne detection (UBHD) \cite{uhom1,uhom2}. Squeezing has been already detected in atomic ensembles, solid-state crystals and waveguides, mechanical oscillators and electrical circuits \cite{sq1,sq2,sq3,sqatom1,sqatom2,sqoptm1,sqoptm2,sqsup}.
BHD requires a single-mode very good shot-noise-limited laser (close to Poissonian photon statistics) with large intensity to directly measure the quadrature of light. Advantageously, average intensity detectors are then sufficient for BHD. In contrast, a low-intensity shot-noise limited laser can be sufficient for UBHD, but the method relies on photon number resolving detectors, which nowadays are not sufficiently accurate. If a process in the matter is pumped by an almost ideal laser light with a fixed known phase locked relative to the probe then, undoubtedly, direct measurement by homodyne detection is very efficient to estimate the squeezing process. 

The problem of the optimal estimation of squeezing originates in the nineties \cite{estsq1}, it is proven that the Heisenberg limit can be achieved \cite{estsq2,estsq12,estsq7,estsq10}. The usual approach is to calculate the quantum Fisher information (QFI) for an arbitrary input state. As in the case of general process tomography \cite{estsq4,estsq6}, the optimal probe should be squeezed, but technically one can provide only a finite level of squeezing and if we take into account the noise, achieving the Heisenberg limit is in general a hard task. Moreover, at a given wavelength a squeezed source is not necessarily achievable. So if our task is to check whether a given process produces squeezing, it is not always realistic to assume to have a squeezed source. Beside that, the achievability of the theoretical maximum of QFI is not a self-evident question, and the optimal measurements are not necessarily realizable with current technology \cite{estsq12}.

We are not trying to push these limits further, we are rather interested in what the minimal requirements of the estimation of squeezing are. Take as an example standard interferometry, where the optimal performance is achieved with a squeezed and displaced probe \cite{estsq13}, but in principle one can perform it even with a thermal source. One can define the interferometric power of an arbitrary process \cite{estsq5}, however, the measure introduced by Adesso is too generic and not developed to estimate the parameters of a specific process. Practical estimation methods usually use homodyne detection, but there is no need to have a strong local oscillator for estimation \cite{Rupp}. There is no recommended estimation method if the conditions for homodyne detection are not fulfilled, namely, if the source of low-intensity radiation is thermal (or close to thermal). In addition, a nonlinear process pumped independently in matter can have an a priori unknown phase; moreover, its phase can fluctuate. Even laser light does not have a well-defined absolute phase and beyond shot-noise-limited lasers it typically contains a considerable amount of thermal noise.

In this paper we verify that classical interferometry with thermal light and limited intensity is sufficient to estimate {\em independent} squeezing processes. In general, it allows estimating all the main phase-independent characteristics of a Gaussian process (i.e., the magnitude of the squeezing, the displacement and the phase-shift). Simultaneously, we analyze the interferometry of a squeezing process with laser light far from the shot-noise-limit. We prove that active interferometry is not required: it does not improve the quality of estimation. We verify the robustness of interferometry if loss and noise are present in the channel. Our results largely relaxed the requirements of existing estimation schemes, thus, they can open many possibilities to estimate nonlinear phase-dependent processes in new materials and soft matter at wavelengths where high-quality laser light and homodyne detection is not available. A possible application could be quantum key distribution with macroscopic states \cite{Usen}, where the source itself provides sufficient enough power to feed the intensity detectors.


\section{Results}

\begin{figure}[!t]
\centering
\includegraphics[width=0.75\columnwidth]{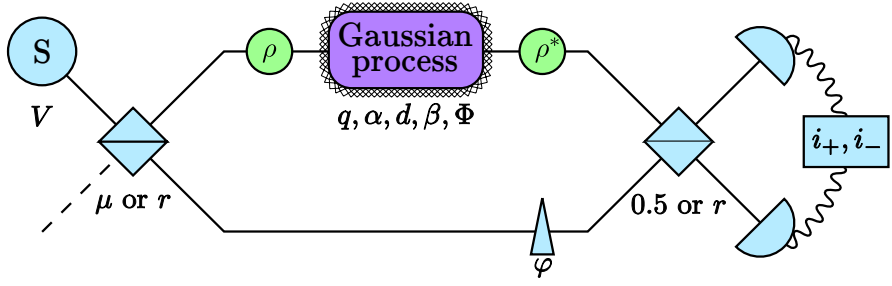}
\caption{\label{scheme} Schematic setup of our model. A Gaussian source is used to feed both the reference and the signal modes (using a BS or OPA), then the signal is modified and finally we perform a homodyne-type measurement.}
\end{figure}

The investigated scheme (see Fig.\ \ref{scheme}) is quite similar to the scheme of standard interferometry using homodyne measurement. The main difference is that the modification of the system is not merely a phase-shift, but an arbitrary Gaussian unitary operator, and the source is not necessarily a coherent local oscillator, it can be an arbitrary state which is not phase-locked to the unknown Gaussian operator. This source is split into two separate modes using an optical element (which can be a beam splitter (BS) or an optical parametric amplifier (OPA)). Then the unknown Gaussian modification, the reconstruction of which is our aim, is performed on the signal state. For that, a joint measurement of the two modes is performed: We apply a phase-shift to the reference (with angle $\varphi$), we couple the two modes on another optical element and use two intensity detectors to obtain $N$ measurement data pairs. 

We decompose the Gaussian unitary process into three parts: 
\begin{equation}\label{channel_form}
\rho^*=D(\gamma)R(\Phi)S(\xi)\rho S(\xi)^\dagger R(\Phi)^\dagger D(\gamma)^\dagger,
\end{equation}
where  $S(\xi)=\exp(1/2 \xi^2 a^{\dagger2}-1/2 \xi^{*2} a^2)$ is the squeezing operator (with $\xi=w \mathrm{e}^{i \alpha}$), $R(\Phi)=\exp(i \Phi a^{\dagger}a)$ is the phase-shift operator, and $D(\gamma)=\exp(\gamma a^{\dagger}-\gamma^* a)$ is the displacement operator (with $\gamma=d \mathrm{e}^{i \beta}$). The squeezing can be described by the magnitude $(q=\mathrm{e}^w)$ and by the direction $(\alpha)$ of the squeezing. The displacement can be described by the magnitude $(d)$ and by the direction $(\beta)$ of the displacement. The combination of these three transformations describes each possible Gaussian transformation of a given Gaussian state which preserves the mixedness of the original state.

The source is independent of the process, meaning their pumping can be physically independent, but more importantly, they can have different frequencies, and thus the phase of the source is not locked to the phase of the process. This means that the phase of the source relative to the phase of the Gaussian process is constantly changing and is unknown in any given moment. So we assume that its phase $\theta$ is random: it is uniformly distributed on $[0,2\pi]$, resulting a rotationally symmetric state in the phase space (e.g., its Wigner-function is bell-shaped for thermal source, toroidal for a coherent source). Let $V$ denote the second moment of quadrature $x$ or $p$ of the random-phase source. Although the obtained formulas in the manuscript are valid for an arbitrary source, due to their practicability, during the numerical investigation we focus mainly on classical Gaussian states: we use a displaced thermal source with standard deviation $R$ and displacement $D$, that is, a Gaussian state with mean value $(D \cos\theta, D \sin\theta)$ and variance $R^2\cdot \mathbb{I}$. Using these sources, we have as a special case thermal states ($D=0$) and coherent states with random phase ($R=1$). That is, for the general Gaussian source we have $\<x^2\>=\<p^2\>=V=R^2+D^2/2$, which can be also derived as a simple function of the mean photon number of the source: $V=2\overline n+1$.

Standard interferometry deals with the estimation of $\Phi$. We are interested also in whether and how accurately parameters $q$ and $d$ can be estimated.

Note that the proposed interferometric scheme is not the simplest one. One could also apply the Gaussian process directly to the source and then measure the output of the process with an intensity measurement. Then the detector would have an average photon number of
\begin{equation}
\<i\>= \frac{\bigg(q^2+\frac{1}{q^2}\bigg)V+d^2}{4}-1,
\end{equation}
and by using two different sources with $V_1$ and $V_2$, the value of $q$ and $d$ could be estimated. However, since we could obtain only two independent equations, if a noise or loss was present their estimation would not be feasible. They would be indistinguishable from the unknown Gaussian process, which would result in a biased estimation of the parameters (and the estimation of $\Phi$ is also not possible without the interferometric setting).


\subsection{Passive interferometry}

In the first case we assume that both optical elements in our setup are beam splitters (BS). The first BS has a reflectivity of $\mu$, while the second is a 50:50 beam-splitter.

With the given parameters, it is easy to calculate the expected difference of the photon numbers:
\begin{equation}\label{minus}
\<i_-\>= \frac{V-1}{2}\sqrt{\mu(1-\mu)}\bigg(q+\frac1q\bigg)\cos(\Phi-\varphi)
\end{equation}
and also the expectancy of the sum of the photon numbers:
\begin{equation}\label{plus}
\<i_+\>= \frac{\bigg(q^2+\frac{1}{q^2}\bigg)V_S+2 V_R+d^2}{4}-1,
\end{equation}
where $V_S=\mu V+1-\mu$ and $V_R=\mu+V-\mu V$ are the variances of the signal and the reference after the first beam splitter.

We assume that the variance of the source ($V$) and the transmittance of the beam splitter ($\mu$) are known, while regarding the parameters of the unknown Gaussian process ($\Phi, q, d$) we have no prior knowledge. Then we can estimate the unknown process by using an unaltered ($\varphi=0$) and an orthogonal ($\varphi=\pi/2$) reference. We have then
\begin{equation}
\<i_-\>_{\varphi=0}= \frac{V-1}{2}\sqrt{\mu(1-\mu)}\bigg(q+\frac1q\bigg)\cos(\Phi),
\end{equation}
\begin{equation}
\<i_-\>_{\varphi=\pi/2}= \frac{V-1}{2}\sqrt{\mu(1-\mu)}\bigg(q+\frac1q\bigg)\sin(\Phi).
\end{equation}

From which we obtain the estimators
\begin{equation}\label{hatphi}
\hat\Phi=\arctan\bigg(\frac{\<i_-\>_{\varphi=\pi/2}}{\<i_-\>_{\varphi=0}}\bigg)
\end{equation}
and
\begin{equation}\label{hatq}
\hat q=\frac{c+\sqrt{c^2-4}}{2},
\end{equation}
where
\begin{equation}
c=2\frac{\sqrt{\<i_-\>_{\varphi=\pi/2}^2+\<i_-\>_{\varphi=0}^2}}{\sqrt{\mu(1-\mu)}(V-1)}.
\end{equation}

Using (\ref{plus}) we obtain the estimator
\begin{equation}\label{hatd}
\hat d=\sqrt{4\<i_+\>+4-\bigg(\hat q^2+\frac{1}{\hat q^2}\bigg)V_S-2 V_R}.
\end{equation}

Let us note that the described method is a straightforward extension of standard interferometry. Knowing the phase of the source is not necessary, it can be anything, the key issue is that for each signal-reference pair there should be a strong correlation in phase (since they came from the same source before they were split by the optical element). The phase-shift is determined the same way as in the literature (checking the angle of the interference pattern), the magnitude of the displacement and the squeezing can be expressed by investigating also the magnitude of the photocurrents beside their angles. 

The angles $\alpha$ and $\beta$ are inaccessible in this scheme since the whole process is phase-insensitive. But that also means that there is no error caused by the phase-instability of the process or the source, which can be problematic in many cases (e.g, when using homodyne measurement and reconstructing the whole covariance matrix). 

We can investigate how these estimators depend on the other parameters ($R$, $D$, $N$, $\mu$ or $r$), however, the figure of merit to use is not self-evident. The quantum Fisher information is a popular choice, however, since we have a minimalistic approach, we are far from its limit. The Fisher information associated with the given measurements would be better suited, however, we can calculate only the value for the normal approximation. The situation is similar if we want to calculate the variance of the estimators, moreover, the estimators are not unbiased, only asymptotically unbiased. So they are not optimal, since for a limited number of measurements either quantity will be imprecise. On other hand, for large values of $N$ the approximation errors will be small, therefore they could be used for numerical optimization, and we can observe variances that are very close to the lower bound obtained from the Cramér-Rao inequality (for detailed calculations and analysis, please see the supplementary material, Sec. 1). 

\begin{figure}[!t]
\centering
\includegraphics[width=0.9\columnwidth]{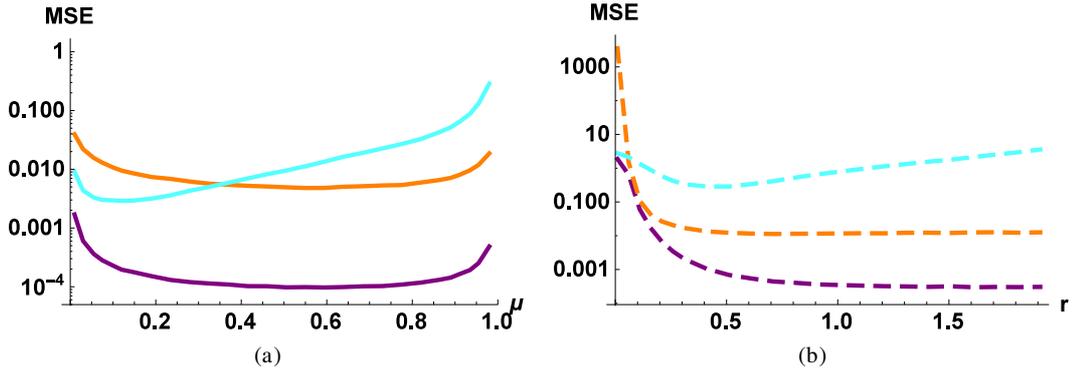}
\caption{\label{ideal1}  MSE of the process estimation as a function of the parameter of optical elements for (a) beam splitters and (b) optical parametric amplifiers. Cyan (light) lines correspond to the estimation of displacement parameter $d$, orange (medium) to squeezing parameter $q$ and purple (dark) to phase-shift parameter $\Phi$. Curves corresponding to beam splitters are solid, results for OPAs are dashed lines. We have parameters $D=10$, $R=5$, $q=1.23$, $\Phi=0.63$, $d=1.67$ and $N=10^4$.}
\end{figure}

To avoid the above problems, we used the empirical mean squared error (MSE):
\begin{equation}
MSE(\hat{q})=\frac{1}{M}\sum_{k=1}^M(\hat{q}_k-q)^2,
\end{equation}
That is, in order to obtain the empirical mean squared error of the estimator based on N measurements, we simulated such a block of N measurements M times (we used $M=10^4$) to draw the required statistic from the M numerically estimated values ($\hat{q}_k$).

In Fig.\ \ref{ideal1}a (solid lines) one can see that neither the fully transmissive, nor the fully reflective beam splitter is useful. The optimum is in-between the two extremes, for the estimation of the displacement a small $\mu$ is optimal, that is, a strong reference and a relatively weak signal (as in the case of homodyne measurement). While for the estimation of the phase-shift or the squeezing the dependence is not as strong, a large $\mu$ (i.e., strong signal) ensures a better performance.

\begin{figure}[!t]
\centering
\includegraphics[width=0.9\columnwidth]{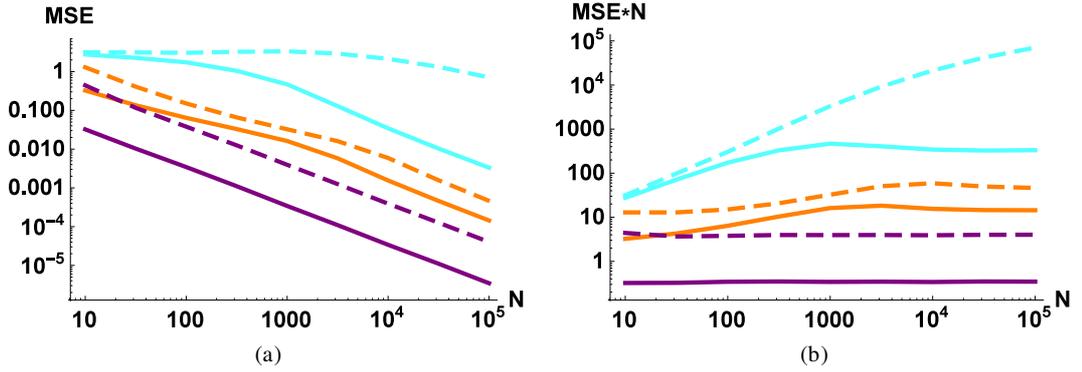}
\caption{\label{ideal0} (a) MSE and (b) MSE$\cdot N$ of the process estimation as a function of the number of measurements ($N$). Cyan (light) lines correspond to the estimation of displacement parameter $d$, orange (medium) to squeezing parameter $q$ and purple (dark) to phase-shift parameter $\Phi$. Curves corresponding to beam splitters are solid, results for OPAs are dashed lines. We have parameters $D=10$, $R=5$, $q=1.23$, $\Phi=0.63$, $d=1.67$, $\mu=0.3$, $r=0.5$.}
\end{figure}

Fig.\ \ref{ideal0}a shows that by using more measurements we can estimate all parameters with an arbitrary precision. More precisely, the speed of convergence is $MSE\sim 1/N$ (Fig.\ \ref{ideal0}b). Let us note that from the theory of interferometry this scaling is the best that we can expect. To achieve better efficiency (like the Heisenberg scaling \cite{estsq2,estsq12,estsq7,estsq10} with $MSE\sim 1/N^2$) one should use even in the phase sensitive case more sophisticated resources (e.g., squeezed vacuum or squeezed coherent states instead of vacuum or coherent states). Since our aim is to have a scheme as simple as possible, we are not investigating these types of extensions in the current manuscript.

\begin{figure}[!t]
\centering
\includegraphics[width=0.9\columnwidth]{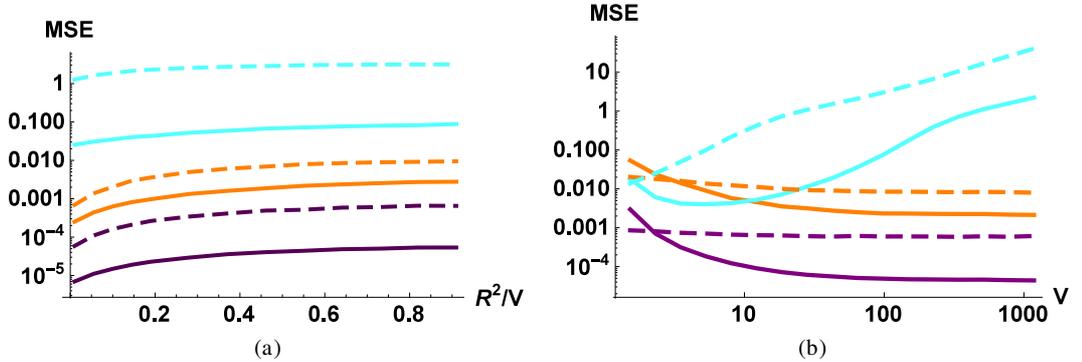}
\caption{\label{ideal2} MSE of the process estimation as a function of the (a) thermal ratio ($R^2/V$) and (b) energy of the source (V). Cyan (light) lines correspond to the estimation of displacement parameter $d$, orange (medium) to squeezing parameter $q$ and purple (dark) to phase-shift parameter $\Phi$. Curves corresponding to beam splitters are solid, results for OPAs are dashed lines. We have parameters $N=10^4$, $\mu=0.3$, $r=0.5$, $q=1.23$, $\Phi=0.63$, $d=1.67$ for (a) $V=100$ and for (b) $R=D$.}
\end{figure}

We can see (Fig.\ \ref{ideal2}a) that for a fixed mean photon number of the source, the source with the minimal thermal part (i.e. a coherent state) gives the minimal error. It could be expected that the coherent source would outperform the thermal source, but the difference is not so large: the variance is only 3-10 times larger in the thermal case. So we can have decent estimates even with a thermal source, which is interesting since usually the estimation is based on classical displacement and the quantum fluctuation is considered to be noise. If the variance of the source is greater, we see (Fig.\ \ref{ideal2}b) different behavior for different parameters: for the squeezing and the phase-shift parameter we can get a more precise estimation, while for the displacement worse. That is, a big thermal state in one case acts as a useful resource, in another more like noise. This is not surprising since the effect of squeezing and phase-shift is much more visible in the interferometric pattern using a strong source, so it outweighs the negative effects of the less imprecise measurement. The increment in energy (photon number) for a given displacement is independent of the source power, so by increasing the source power the signal remains constant, but the noise increases.


\subsection{Active interferometry}

Previously we considered only passive optical elements, here we investigate the case when one or both of the beam-splitters are changed to an optical parametric amplifier (OPA). Let us assume that the OPA is fed by a pump, which results in a two-mode squeezing with parameters $r$ and $\phi$. We assume that as the source, the angle of this two-mode squeezing is not locked to the unknown Gaussian process either. Therefore we can assume that $\phi$ is also random.

An immediate consequence is that if we exchange only one BS for an OPA, then we do not see any interference pattern, since the randomness of $\phi$ cancels it out completely. However, if we have two OPAs which have the same random $\phi$ parameter (that is, they are phase-locked to each other, but not to the unknown process or to the source), then we can perform a similar analysis as in the BS case.

Once again, we can calculate the expected difference in the photon numbers:
\begin{equation}\label{minus_TWB}
\<i_-\>=\frac{\bigg(q^2+\frac{1}{q^2}\bigg)V_S+d^2-2 V_R}{4},
\end{equation}
and also the sum of the photon numbers:
\begin{equation}\label{plus_TWB}
\<i_+\>=\cosh (2r_2) \frac{\bigg(q^2+\frac{1}{q^2}\bigg)V_S+d^2+2 V_R}{4}+\sinh(2r_2)\sinh(2r_1) \frac{V+1}{4}\bigg(q+\frac1q\bigg)\cos(\Phi+\varphi)-1
\end{equation}
where $r_1$ and $r_2$ are the parameters of the first and second OPA, respectively, and $V_S=\cosh(r_1) V+\sinh(r_1)$ and $V_R=\sinh(r_1) V+\cosh(r_1)$ are the variances of the signal and the reference after the first beam splitter. For simplicity, in the further discussion we will always assume that the two OPAs have the same parameter: $r_1=r_2=:r$.

We can see that these equations are different from those obtained in the beam splitter case (e.g., the interfering pattern is moved from the difference to the sum), but structurally consist of the same parts as Eq.\ (\ref{minus}) and (\ref{plus}). Therefore, we can obtain estimators for $\Phi$, $q$ and $d$ in a similar way as for passive interferometry.

Looking at the estimation efficiency (Fig.\ \ref{ideal1}-\ref{ideal2}, dashed lines), we can conclude that most of the conclusions drawn for BS are also valid for OPAs. The coupling effect ($r$) should be neither too small nor too large; we have $1/N$ scaling; the coherent source is only slightly better than a thermal source; a stronger source improves the estimation of the phase-shift and the squeezing, but decreases the efficiency of the estimation of the modulation. The main difference is that for the investigated parameters passive interferometry in most cases outperforms the active counterpart (the exception is when the source is very weak, i.e., close to a vacuum). At first it may sound counterintuitive that using an outside source of energy (as a pump of OPA) to produce entanglement results in a worse estimation. But for example in the case of the phase-shift, it is known that the Fisher information using a BS is larger than by using an OPA \cite{estsq13}. The OPA can outperform BS around the optimal working point, that is, when the angle of input state (and the angle of OPA pump) is optimal. In the phase-insensitive case, we have the average efficiency, so actually it is not surprising that passive interferometry is more efficient. Let us also note that in practice the parameter range for $r$ is limited, achieving the optimal working point (which is around $r=0.5$, Fig.\ \ref{ideal1}b) is technically not as trivial as for the beam-splitter.


\subsection{The effect of loss and noise}

So far, we have discussed the case of an ideal process containing neither noise nor loss. In the following, we will discuss how different errors influence the results of the estimation.

First, let us note that we are using intensity detectors. These do not work perfectly, but if we can estimate their quantum efficiency, then using that we can obtain an unbiased estimate of mean photon numbers. We investigate two typical cases with channel-type and process-type errors. The first one appears between the preparation and the measurement phases to both the signal and the reference mode, while the latter during the implementation of the Gaussian process. Naturally, in the presence of loss and noise the previously obtained estimators are biased, but by using an additional measurement 
round the effect of the errors can be estimated in either case. 

\begin{figure}[!h]
\centering
\includegraphics[width=0.9\columnwidth]{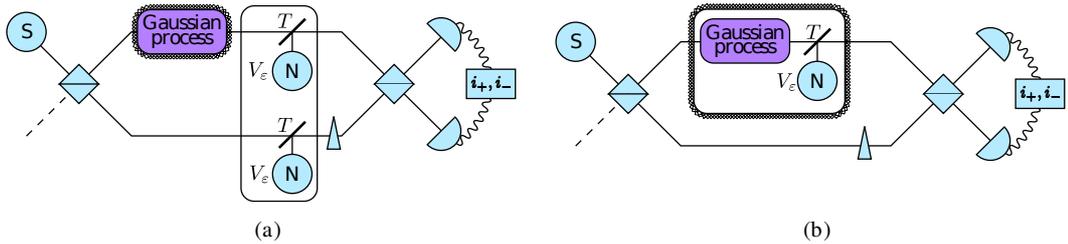}
\caption{\label{scheme2} Schematic setup for (a) channel-type errors and (b) process-type errors.}
\end{figure}

In the case of channel loss (see Fig. \ref{scheme2}a) we can perform first a calibration round without applying the process. Then the only unknown parameters will be the noise and loss of the channel, hence we can estimate both using intensity measurements. And in the second step we perform a regular, noisy setup, and we can determine the parameters of the Gaussian process by modifying the estimators (\ref{hatphi})-(\ref{hatd}) to include the  errors (which we estimated in the first step). 

The case of process-type errors (see Fig. \ref{scheme2}b) is a little more complicated because we can not estimate the errors independently of the process. However, we can use sources with different energies ($V_1$ and $V_2$), which will give use twice as many equations on the intensities. The total intensity $\<i_+\>$ will include the same noise independently of $V_i$, so by using the difference of the intensities for the two sources $(\<i_+\>_{V_2}-\<i_+\>_{V_1})$ we can eliminate the process noise. Using this additional equation we can already estimate the process loss and by that we can estimate the phase-shift and the squeezing, but not the displacement (for detailed calculations and analysis for both cases, please see the supplementary material, Sec. 2). 

It is no surprise that it is possible to estimate the phase-shift, we know from the theory of standard interferometry that it can be done even in the presence of errors. The estimation of the displacement can be a problem, since in a phase-insensitive setup it acts exactly as noise (both are only present in the increase of energy but not in the interference), so with limited measurement possibilities, it can happen that they are indistinguishable. However, the estimation of squeezing presents no difficulties, the difference is that the squeezing also changes the difference of photon numbers (that is, it has a visible effect in the interference), not just the sum of photon numbers (i.e., in the total energy).

\begin{figure}[!t]
\centering
\includegraphics[width=0.75\columnwidth]{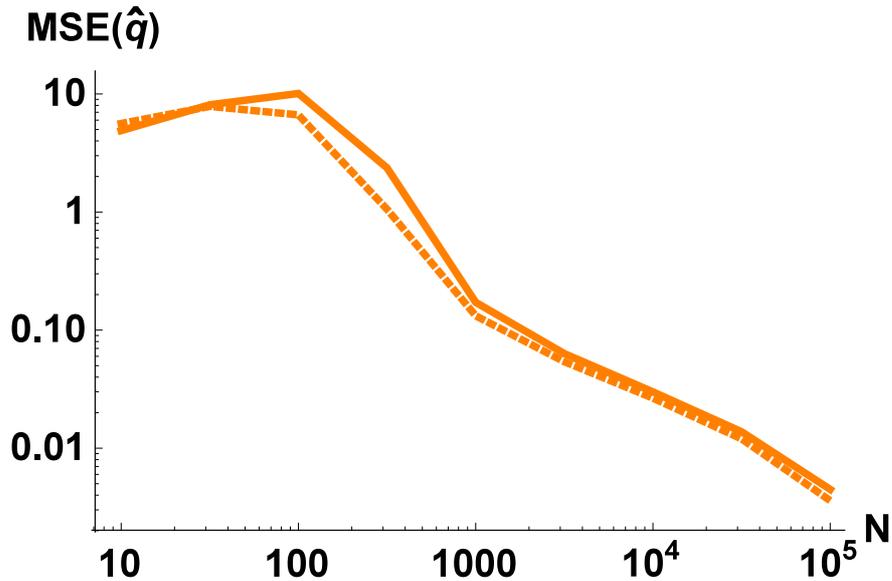}
\caption{\label{both_q} MSE of the estimation of squeezing ($\hat{q}$) as a function of $N$ for BS using thermal sources with variances $V_1=10$ (solid line) and  $V_1=1000$ (dashed line). We assume that during the process both process-type ($T_1=0.9$, $V_{\eps_1}=1.1$) and channel-type ($T_2=0.7$, $V_{\eps_2}=1.3$) errors are present simultaneously. We have parameters $V_2=4V_1$, $q=1.23$, $\Phi=0.63$, $d=1.67$, $\mu=0.3$.}
\end{figure}

Moreover, we can estimate the magnitude of the squeezing with both channel and process-type noise present only by using weak thermal sources (Fig.\ \ref{both_q}). For that we have to combine the two estimation methods discussed above. Note that doing this we cannot estimate the noises and the displacement of the process, but those are not required to estimate the squeezing. 

Unlike the ideal case, when $V=1000$ had a $2-3$ times lower variance than  $V=10$ (see Fig.\ \ref{ideal2}b), here the quality of the estimation does not depend significantly on the strength of the source, $V=10$ and $V=1000$ are barely distinguishable. For larger states the effect of squeezing will be more visible, but also the effects of errors will be larger and seemingly these two effects cancel each other out quite well (the $V=1000$ case is still better, but only with 10-30\%). If we compare the values of MSE with the noiseless version, we can see that the estimation efficiency is worse if additional noise is present (as it is expected). Nevertheless, the characteristic of the lines is the same, both converge to zero with an error of the order of $1/N$. That is, with a sufficient number of measurements we can estimate the magnitude of squeezing with arbitrary precision even with a weak thermal source and with different kinds of noises and losses present.


\section{Discussion}

In our work we investigated phase-insensitive measurements to obtain the estimators for an unknown general Gaussian unitary process. The common method for estimating only the phase-shifts is the well-known interferometry. We extended the standard method of interferometry to estimate squeezing and displacement without using more sophisticated tools (e.g., homodyne measurements).

The efficiency of the estimation depends on numerous details. In general we can conclude that the estimation of the phase-shift is the most accurate, while the estimation of the displacement is the worst. This is not surprising since the phase-shift comes immediately from the interference pattern, while the effect of the displacement is only visible as an increase in the mean photon number. Our method can be applied to both coherent and thermal sources, and while the former yields more precise estimates the difference is not substantial. We examined the effect of including either beam splitters or optical parametric amplifiers in the interferometric scheme and constructed the estimators for both cases. We observed very similar behavior, with a slight advantage of passive elements.

We also investigated the effect of loss and noise. In general, we can say that the estimators are not robust against these errors: only the phase-shift can be estimated well, the estimation of squeezing and modulation will be biased even for a relatively low level of losses or noise. However, if we know the structure of errors, we can estimate those too, and by incorporating these estimators we can already estimate the unknown process quite well. The only problem arises when we cannot estimate the error and the process independently, since then the noise and the displacement have the same effect due to the phase-insensitivity, hence they are indistinguishable. In spite of the displacement, the effect of squeezing is clearly visible in the interference pattern, so we can conclude that using our method it is possible to estimate the squeezing as well as the phase-shift in all possible scenarios very well even with both types of discussed errors present and weak thermal sources.


\section*{Acknowledgements}

L.R and R.F. acknowledges project GB14-36681G of the Czech Science Foundation.

\pagebreak

{\huge \bf Supplementary part}

\section*{Approximation of Fisher information and mean squared error}

The estimators of the unknown parameters are non-linear, therefore their quality (FI or MSE) cannot be calculated analytically. In the following we will describe an approximation for the estimator of the squeezing parameter $q$ for a thermal source and passive optical elements (beam splitters). For different setups (i.e., using a displaced thermal source, active optical elements or different estimators) the calculations can be performed similarly, but we refrain from it here because the conclusion would be the same.

Let us remind ourselves that the estimator of the squeezing magnitude is
\begin{equation}
\hat q=\frac{c+\sqrt{c^2-4}}{2},
\end{equation}
where
\begin{equation}
c=2\frac{\sqrt{\<i_-\>_{\varphi=\pi/2}^2+\<i_-\>_{\varphi=0}^2}}{\sqrt{\mu(1-\mu)}(V-1)}.
\end{equation}

\subsection*{Mean squared error}

The MSE cannot be calculated analytically since $\<i_-\>$ is already complicated enough so that its likelihood function can be problematic to give. Moreover, we need to perform non-linear transformations to obtain the value of $\hat{q}$, so to calculate $MSE(\hat{q})$ we need to apply some approximation. In the main text we used numerical simulation and empirical MSE, which has the advantage that with a large enough iteration number ($M$) we can get arbitrarily close to the real MSE. Here we will approximate the distribution with a normal distribution with the same mean and variance. As we will see it results in a good approximation for large numbers of measurements ($N$) and since it is a closed formula it can be calculated instantaneously. However, it can be imprecise for small values of $N$. 

The properties of $c$ can be calculated straightforwardly:
\begin{equation}
m:=\bE (c)=q+\frac{1}{q}
\end{equation}
and
\begin{equation}
\sigma^2:=\Var(c)=\frac{1}{N}\bigg(q^2+\frac{1}{q^2}\bigg)\bigg(2+\frac{V}{(1-\mu)\mu(V-1)^2}\bigg)
\end{equation}

The value of $c^2$ can be calculated using this, since if we assume $c$ is normally distributed, then $c^2$ will be non-central $\chi^2$ distributed. While $c^2-4$ is only displaced by the mean of $-4$, so we obtain
\begin{equation}
\bE (c^2-4)=m^2+\sigma^2-4
\end{equation}
and
\begin{equation}
\Var(c^2-4)=2 \sigma^2(2 m^2+\sigma^2).
\end{equation}

Once again $c^2-4$ is not normally distributed, but it can be approximated (for a certain regime) with a normal distribution. Then its mean can be calculated using Tricomi's confluent hypergeometric U-function:
\begin{equation}
\bE \sqrt{c^2-4}=\frac{1-i}{2^{1/4}}\cdot \Var^{1/4}(c^2-4)\cdot U\bigg(-\frac{1}{4},\frac{1}{2},-\frac{\bE^2 (c^2-4)}{2 \Var(c^2-4)}\bigg),
\end{equation}
and the variance is
\begin{equation}
\Var \sqrt{c^2-4}=\bE (c^2-4)-\bE^2 \sqrt{c^2-4}.
\end{equation}

Using the above it is already easy to obtain the properties of $\hat{q}$:
\begin{equation}\label{eeq}
\bE (\hat{q})=\frac{\bE (c)+\bE \sqrt{c^2-4}}{2}
\end{equation}
and
\begin{equation}\label{varq}
\Var (\hat{q})=\frac{\Var (c)+\Var \sqrt{c^2-4}}{2}
\end{equation}

From (\ref{eeq}) it is visible that $\hat{q}$ is not an unbiased estimator, numerically one can check that it is still asymptotically unbiased. If the bias is small and the normal approximation is valid, then (\ref{varq}) will be close to the actual MSE of $\hat{q}$. 

\subsection*{Fisher information}

The situation is similar in the case of Fisher information. Let us derive the maximally available information on $q$ from $\<i_-\>$. For that we approximate $\<i_-\>$ with a normal distribution.
Its mean is 
\begin{equation}
\bE (\<i_-\>)=\frac{V-1}{2}\sqrt{\mu(1-\mu)}\bigg(q+\frac1q\bigg)\cos(\Phi-\varphi).
\end{equation}
The last term is a constant multiplier, so let us use $\varphi=\Phi$ to obtain the maximal FI.
In this case the variance will be
\begin{equation}
\Var (\<i_-\>)=\frac{1}{4N}\bigg(q^2+\frac{1}{q^2}\bigg)\bigg(2(1-T)T(V-1)^2+V\bigg).
\end{equation}

Let $f(x,q)$ be the likelihood function of a normal distribution with the same mean and variance. Then the Fisher information can be calculated as 
\begin{equation}
I_{normal}=\int \bigg(\frac{\partial \log f(x,q)}{\partial q}\bigg) f(x,q) dx.
\end{equation}
The derivative of the likelihood function can be expressed explicitly and can be rewritten in the following form:
\begin{equation}
I_{normal}=\int p_4(x) e^{-a(x-b)^2} dx,
\end{equation}
where $p_4(x)$ is a fourth order polynomial, $a$ and $b$ are functions of the parameters. Since $\int_{-\infty}^{\infty} x^k e^{-a(x-b)^2} dx$ can be expressed analytically for every value of $k$, in principle we can calculate the Fisher information analytically, however, its formula will be too extensive to include here (nevertheless, for a computer it is not a problem). 

The Fisher information is related to the variance through the Cram\'er-Rao inequality:
\begin{equation}
\Var(\hat{q})\ge I^{-1},
\end{equation}
that is, the inverse of the Fisher information is a lower bound for the variance, and so also for the MSE. However, it is unclear whether we can get close to this bound or not. Another issue is that we have only an approximation of the Fisher information; where the normal approximation does not hold, we do not know how the actual Fisher information and $I_{normal}$ are related.

\subsection*{Numerical results}

\begin{figure}[!t]
\centering
\begin{tabular}{cc}
\includegraphics[width=0.42\columnwidth]{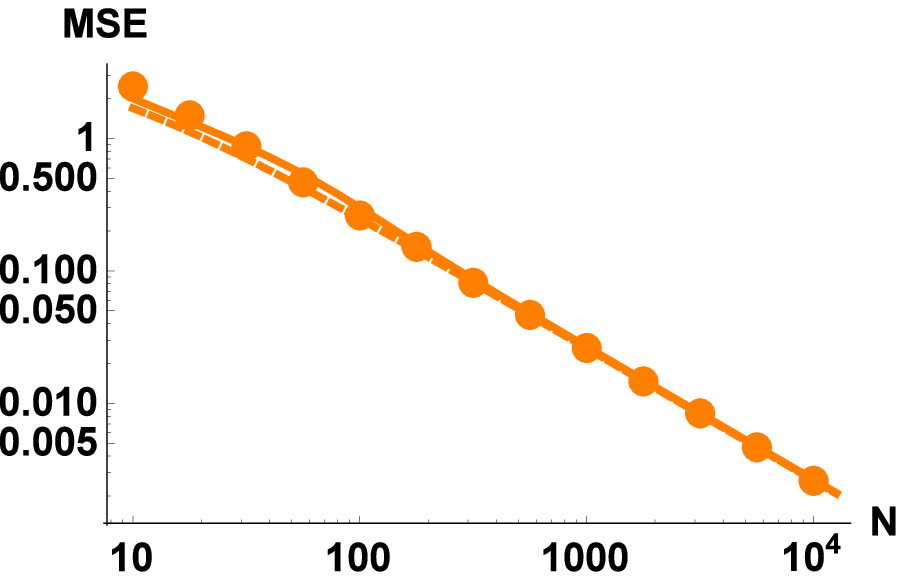}&
\includegraphics[width=0.42\columnwidth]{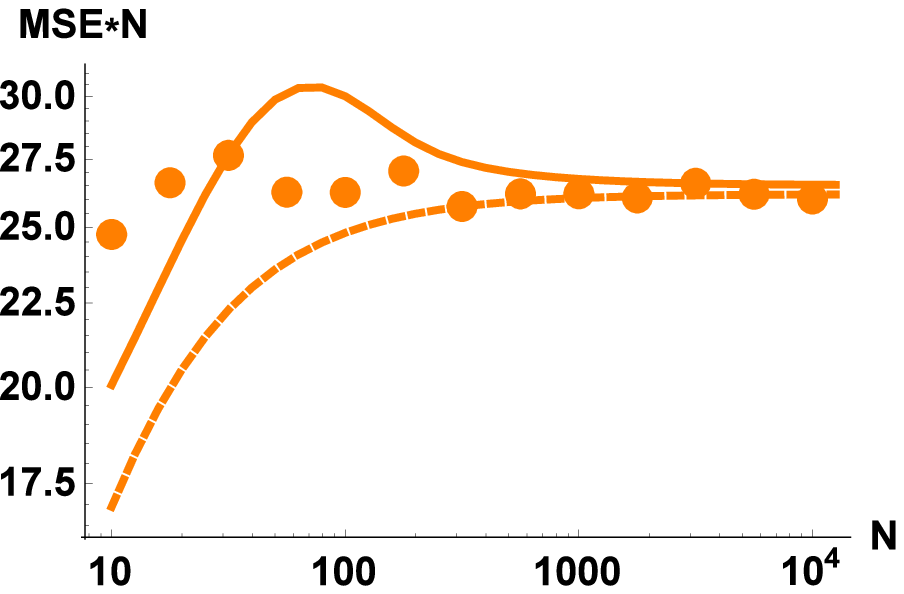}
\\(a)&(b)\end{tabular}
\caption{\label{approx} MSE (a) and MSE$\cdot N$ (b) of the estimation of squeezing parameter $q$ with different approximations: dots correspond to the numerical simulation, solid lines to the approximated variance, dashed lines come from Cram\'er-Rao inequality using the approximated Fisher information. We have parameters $R=5$, $D=0$, $T=0.2$ and $q=3$.}
\end{figure}

The plots of a particular case can be seen in Fig. \ref{approx}. In the left subfigure we can see that the three methods produce very similar errors. The MSE is of order $1/N$, so we plotted the normalized version too (right subfigure). Here, we can see better that for large values of $N$ the values from the numerical simulation almost coincide with the approximation of the variance and the Fisher information. This is expected since according to the central limit theorem the given distributions indeed converge to a normal distribution. So for large values of $N$ the approximation will be really good. On the other hand, below $N=1000$ the 3 curves separate. The magnitude of the differences in the current settings goes up to 50\%, which is in some situations significant. So we used the numerical simulation even though its calculation is the slowest and does not give smooth curves, just because it is working in all regimes, while the other approximations can give us systematically wrong results.


\section*{The effect of loss and noise}

\subsection*{Channel-type errors}

First, we will investigate the effect of errors by applying a channel between the preparation and the measurement phases to both the signal and the reference mode. The channel is modeled by a beam-splitter coupled to a thermal source (Fig.\ \ref{scheme2} left subfigure). The transmittance of the beam splitter ($T$) defines the loss of the channel, the variance of the auxiliary mode ($V_\eps$) defines the noise of the channel.

\begin{figure}[!t]
\centering
\begin{tabular}{cc}
\includegraphics[width=0.42\columnwidth]{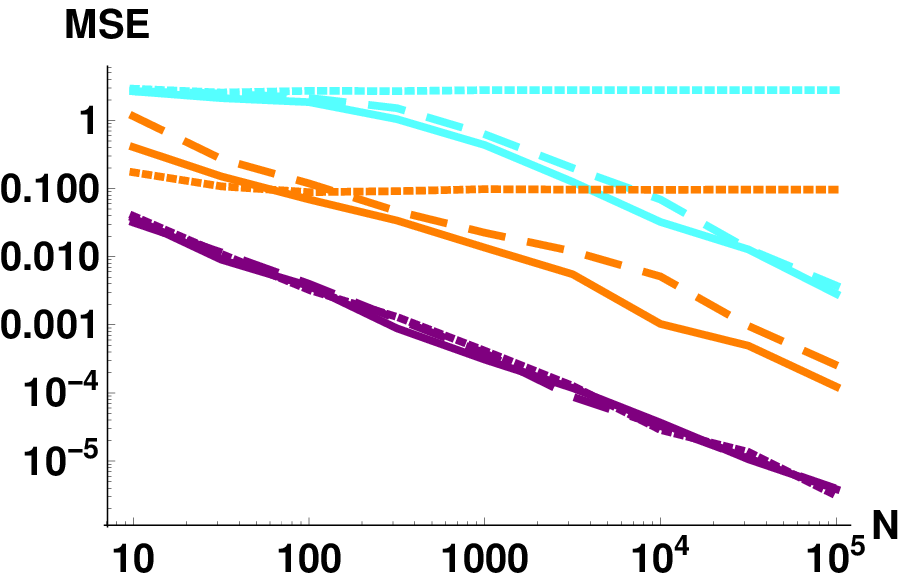}&
\includegraphics[width=0.42\columnwidth]{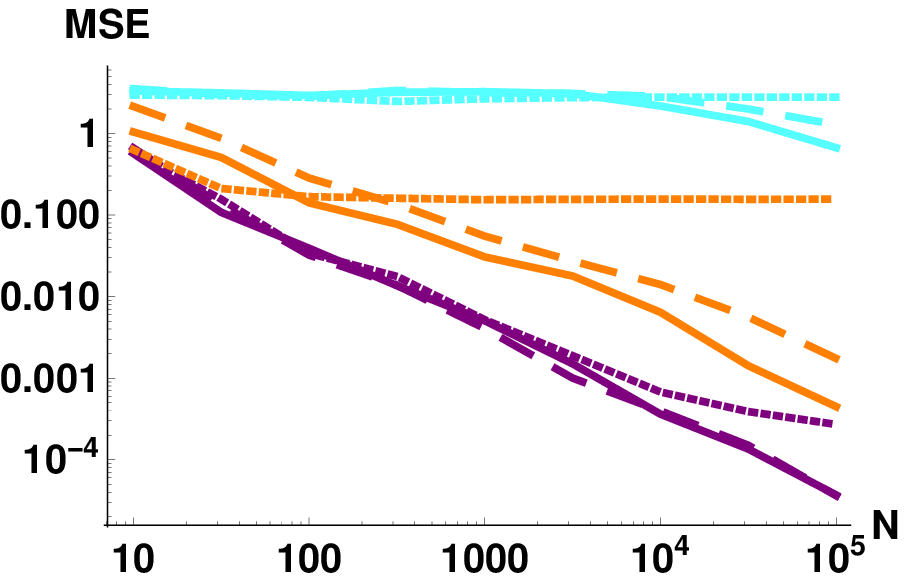}
\\(a)&(b)\end{tabular}
\caption{\label{bias} MSE of the process estimation with channel-type error as a function of $N$ (a) for BS and (b) for OPA. Cyan (light) lines correspond to the estimation of the displacement $d$, orange (medium) to the squeezing $q$ and purple (dark) to the phase-shift $\Phi$. We plotted the cases when we know the noise a priori (solid lines), we estimate the noise (dashed lines), we assume that the channel is ideal (dotted lines). We have parameters $V=75$, $q=1.23$, $\Phi=0.63$, $d=1.67$, $T=0.9$, $V_\eps=1.1$, for (a) $\mu=0.3$ and for (b) $r=0.5$.}
\end{figure}

If we use the estimators defined in the main part the estimation of the parameters can be biased. Indeed, by increasing the number of measurements the estimation cannot converge to the real values of the parameters of the unknown process (Fig.\ \ref{bias}, dotted lines). Therefore the estimators obtained in the noiseless case have limited applicability in a realistic setting; we should take into account the effect of the channel as well (once again, we will discuss in detail only the BS case, the OPA case can be handled using the same steps with similar calculations).

If we include also the noise and the loss in our calculations, we have for the expected difference in the photon numbers:
\begin{equation}\label{minus_noise}
\<i_-\>= T \cdot \frac{V-1}{2}\sqrt{\mu(1-\mu)}\bigg(q+\frac1q\bigg)\cos(\Phi-\varphi)
\end{equation}
and for the sum of the photon numbers:
\begin{equation}\label{plus_noise}
\<i_+\>= T\cdot \frac{\bigg(q^2+\frac{1}{q^2}\bigg)V_S+2 V_R+d^2}{4}+V_\eps(1-T)-1.
\end{equation}

The simplest way to estimate the parameters of the channel is to calibrate the setup prior to the actual measurements. In the first step, we do not modify the signal (i.e., $\rho^*=\rho$), and we let the unaltered signal and the reference pass through the channel. 

In mathematical terms that means that we should substitute into (\ref{minus_noise})-(\ref{plus_noise}) the values $q=1, d=\Phi=\varphi=0$:
\begin{equation}
\<i_-\>_{\textrm{calibr.}}= T \cdot (V-1)\sqrt{\mu(1-\mu)}
\end{equation}
and
\begin{equation}
\<i_+\>_{\textrm{calibr.}}= T\cdot \frac{V}{2}+V_\eps(1-T)-1;
\end{equation}
from which we can estimate the value of the loss
\begin{equation}
\hat T=\frac{\<i_-\>_{\textrm{calibr.}}}{(V-1)\sqrt{\mu(1-\mu)}}.
\end{equation}
and the noise
\begin{equation}
\hat V_\eps=\frac{\<i_+\>_{\textrm{calibr.}}+1-\hat T(V+1)/2}{1-\hat T}.
\end{equation}
Using these we can get unbiased estimates of the parameters $T$ and $V_\eps$.

Now by knowing the channel parameters, in the second step we can solve Eq.\ (\ref{minus_noise})-(\ref{plus_noise}) similarly as in the ideal case, obtaining estimators for the unknown parameters $\Phi$, $q$ and $d$ (Fig.\ \ref{bias}, dashed lines). We can see that these estimators are almost as efficient as if we knew the channel parameters $T$ and $V_\eps$ precisely (Fig.\ \ref{bias}, solid lines). That is, by using a calibration round channel-type noises can be handled exceptionally well.

\subsection*{Process-type errors}

The second naturally arising error can appear during the implementation of the Gaussian process. This error is also modeled by a beam-splitter coupled to a thermal source (Fig.\ \ref{scheme2} right subfigure), with a loss of $T$ and a noise of $V_\eps$. The difference is that in this case the reference is unaltered, and more importantly, we cannot use a calibration round without applying the process since the error is only present if the unknown Gaussian process is present, too.

\begin{figure}[!ht]
\centering
\begin{tabular}{cc}
\includegraphics[width=0.42\columnwidth]{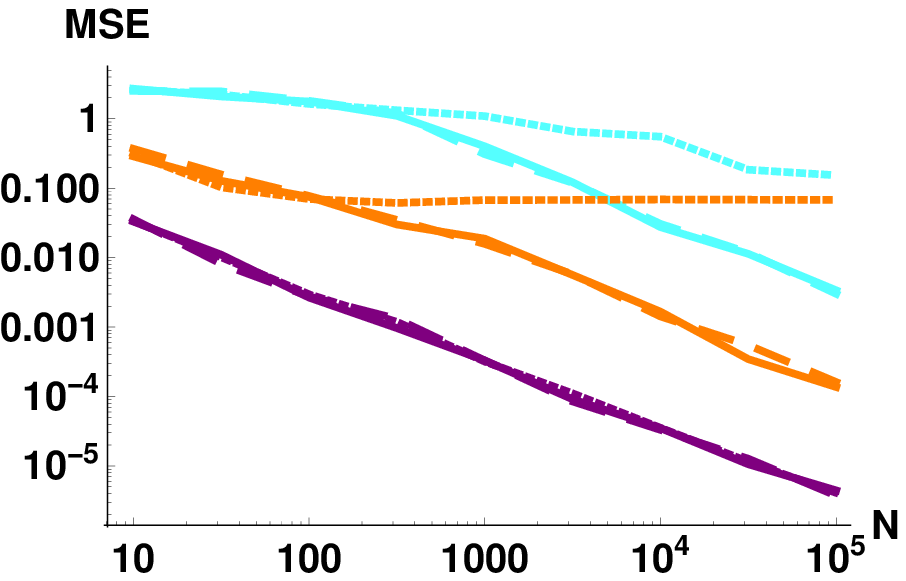}&
\includegraphics[width=0.42\columnwidth]{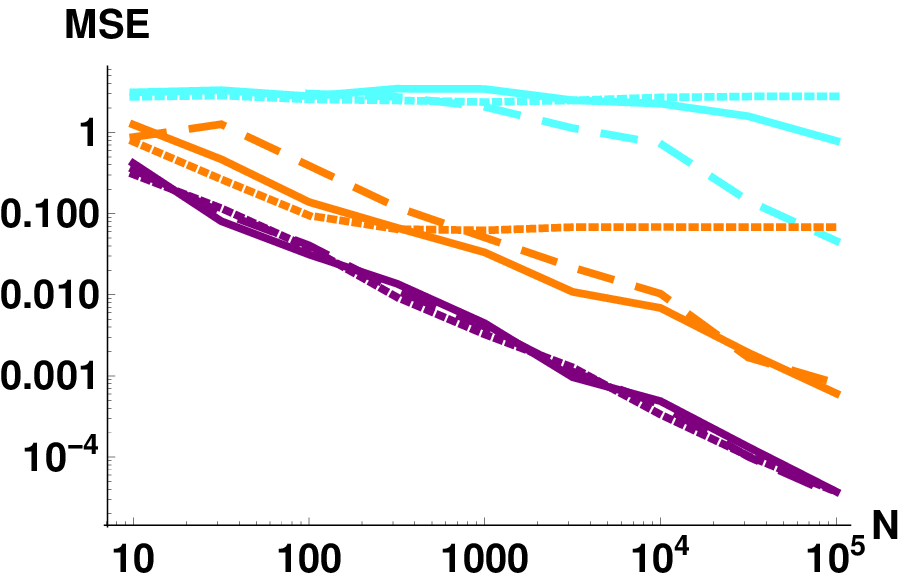}
\\(a)&(b)\end{tabular}
\caption{\label{bias2} MSE of the process estimation with process-type error as a function of $N$ (a) for BS and (b) for OPA. Cyan (light) lines correspond to the estimation of the displacement $d$, orange (medium) to the squeezing $q$ and purple (dark) to the phase-shift $\Phi$. We plotted the cases when we know the noise a priori (solid lines), we estimate the noise (dashed lines), we assume that the channel is ideal (dotted lines). We have parameters $V_1=75$, $V_2=300$, $q=1.23$, $\Phi=0.63$, $d=1.67$, $T=0.9$, $V_\eps=1.1$, for (a) $\mu=0.3$ and for (b) $r=0.5$.}
\end{figure}

If we use the ``naive'' estimators defined for the ideal case, the estimation of the parameters will be biased again (Fig.\ \ref{bias2}, dotted lines). Therefore we have to estimate the errors, but since we cannot estimate the errors in an independent calibration round, we can only do so by repeating the experiment twice using sources with different strengths (i.e., having $V_1$ and $V_2$ instead of a single $V$).

Now for process-type errors instead of Eq. (\ref{minus_noise}) we have
\begin{equation}\label{minus_prep}
\<i_-\>_V= \sqrt{T}\bigg(q+\frac1q\bigg) \cdot \frac{V-1}{2}\sqrt{\mu(1-\mu)}\cos(\Phi-\varphi)
\end{equation}
and instead of Eq. (\ref{plus_noise}) we obtain
\begin{equation}\label{plus_prep}
\<i_+\>_V= T\cdot \frac{\bigg(q^2+\frac{1}{q^2}\bigg)V_S+d^2}{4}+\frac{V_R}{2}+\frac{V_\eps(1-T)}{2}-1.
\end{equation}

We can estimate $\Phi$ the usual way from (\ref{minus_prep}). Using two different values of $V$ in (\ref{plus_prep}) we gain an alternative equation:
\begin{equation}\label{plus_prep_diff}
\<i_+\>_{V_2}-\<i_+\>_{V_1}= T\bigg(q^2+\frac{1}{q^2}\bigg) \cdot \frac{\mu\Delta V}{4}+\frac{(1-\mu)\Delta V}{2},
\end{equation}
with $\Delta V=V_2-V_1$, which combined with (\ref{minus_prep}) provides sufficient information to estimate the parameters $T$ and $q$ beside $\Phi$. However, the effects of the displacement $d$ and the noise $V_\eps$ cannot be distinguished. Neither appears in the interference, they only result in an additional increase in the mean number of photons (i.e., both are seemingly just noise), thus to distinguish them we should use a phase-sensitive measurement. However, if the noise is small and we assume that all of the additional energy is coming from the displacement, we can still have a reasonable estimation precision (Fig.\ \ref{bias2}, dashed lines). The efficiency of these estimators is almost as good as if we knew the error parameter precisely (Fig.\ \ref{bias2}, solid lines).


\begin{thebibliography}{99}

\bibitem{Glauber}
Glauber, R.J. Coherent and incoherent states of the radiation field, \textit{Phys. Rev.} \textbf{131}, 2766 (1963).

\bibitem{Grangier}
Grangier, P., Roger, G. \& Aspect, A. Experimental evidence for a photon anticorrelation effect on a beam splitter: a new light on single-photon interferences, \textit{Europhys. Lett.} \textbf{1}, 173 (1986).

\bibitem{hom}
Hong, C. K., Ou, Z. Y. \& Mandel, L., Measurement of subpicosecond time intervals between two photons by interference
\textit{Phys. Rev. Lett.} \textbf{59}, 2044 (1987).

\bibitem{Jezek}
Je\v zek, M. \textit{et al.} Experimental test of the quantum non-Gaussian character of a heralded single-photon state, \textit{Phys. Rev. Lett.} \textbf{107}, 213602 (2011).

\bibitem{anti1}
Nothaft, M. \textit{et al.} Electrically driven photon antibunching from a single molecule at room temperature,
\textit{Nature Comm.} \textbf{3}, 628 (2011).

\bibitem{anti2}
Wientjes, E., Renger, J., Curto, A. G., Cogdell, R. \& van Hulst, N. F. Strong antenna-enhanced fluorescence of a single light-harvesting complex shows photon antibunching, \textit{Nature Comm.} \textbf{5}, 4236 (2014).

\bibitem{anti3}
Koperski, M. \textit{et al.} Single photon emitters in exfoliated WSe2 structures,
\textit{Nature Nanotech.} \textbf{10}, 503–506 (2015).

\bibitem{anti4}
Ma, X., Hartmann, N. F., Baldwin, J. K. S., Doorn, S. K. \& Htoon, H. Room-temperature single-photon generation from solitary dopants of carbon nanotubes, \textit{Nature Nanotech.} \textbf{10}, 671–675 (2015).

\bibitem{hom1}
Lang, C. \textit{et al.} Correlations, indistinguishability and entanglement in Hong–Ou–Mandel experiments at microwave frequencies, \textit{Nature Phys.} \textbf{9}, 345–348 (2013).

\bibitem{hom2}
Silverstone, J. W. \textit{et al.} On-chip quantum interference between silicon photon-pair sources, \textit{Nature Phot.} \textbf{8}, 104–108 (2014).

\bibitem{hom3}
Gschrey, M. \textit{et al.} Highly indistinguishable photons from deterministic quantum-dot microlenses utilizing three-dimensional in situ electron-beam lithography, \textit{Nature Comm.} \textbf{6}, 7662 (2015).

\bibitem{hom4}
Somaschi, N. \textit{et al.} Near-optimal single-photon sources in the solid state, \textit{Nature Phot.} \textbf{10}, 340–345 (2016).

\bibitem{sqexp1}
Breitenbach, G., Schiller,  S. \& Mlynek, J. Measurement of the quantum states of squeezed light, \textit{Nature} \textbf{387}, 471–475 (1997).

\bibitem{sqexp2}
Andersen, U. L., Gehring, T., Marquardt, Ch. \& Leuchs, G. 30 years of squeezed light generation, 
\textit{Phys. Scr.} \textbf{91}, 053001 (2016).

\bibitem{BW}
Born, M. \& Wolf,  E. \textit{Principles of Optics: Electromagnetic Theory of Propagation, Interference and Diffraction of Light} (Cambridge University Press, 1999).

\bibitem{MW}
Mandel, L. \& Wolf,  E. \textit{Optical Coherence and Quantum Optics} (Cambridge University Press, 1995).

\bibitem{bhom1}
Eberle, T. \textit{et al.} Quantum enhancement of the zero-area Sagnac interferometer topology for gravitational wave detection, \textit{Phys. Rev. Lett.} \textbf{104}, 251102 (2010).

\bibitem{bhom2}
Mehmet, M. \textit{et al.} Squeezed light at 1550 nm with a quantum noise reduction of 12.3 dB, \textit{Opt. Exp.} \textbf{19}, 25763 (2011).

\bibitem{uhom1}
Laiho, K., Cassemiro, K. N., Gross, D. \& Silberhorn, Ch. Probing the negative Wigner function of a pulsed single photon point by point, \textit{Phys. Rev. Lett.} \textbf{105}, 253603 (2010).

\bibitem{uhom2}
Harder, G. \textit{et al.} Local sampling of the Wigner function at telecom wavelength with loss-tolerant detection of photon statistics, \textit{Phys. Rev. Lett.} \textbf{116}, 133601 (2016).

\bibitem{sq1}
Slusher, R., Hollberg, L., Yurke, B., Mertz, J. \& Valley, J. Observation of squeezed states generated by four-wave mixing in an optical cavity, \textit{Phys. Rev. Lett.} \textbf{55}, 2409 (1985).

\bibitem{sq2}
Wu, L. A., Kimble, H. J., Hall, J. L. \& Wu, H. Generation of squeezed states by parametric down conversion, \textit{Phys. Rev. Lett.} \textbf{57}, 2520 (1986).

\bibitem{sq3}
Shelby, R. M., Levenson, M. D., Walls, D. F., Aspect, A. \& Milburn, G. J. Generation of squeezed states of light with a fiber-optic ring interferometer, \textit{Phys. Rev. A} \textbf{33}, 4008 (1986).

\bibitem{sqatom1}
Ourjoumtsev, A. \textit{et al.} Observation of squeezed light from one atom excited with two photons, \textit{Nature} \textbf{474}, 623–626 (2011).

\bibitem{sqatom2}
Schulte, C. H. H. \textit{et al.} Quadrature squeezed photons from a two-level system, \textit{Nature} \textbf{525}, 222 (2015).

\bibitem{sqoptm1}
Safavi-Naeini, A. H. \textit{et al.} Squeezed light from a silicon micromechanical resonator, \textit{Nature} \textbf{500}, 185–189 (2013).

\bibitem{sqoptm2}
Wollman, E. E. \textit{et al.} Quantum squeezing of motion in a mechanical resonator, \textit{Science} \textbf{349}, 952 (2015).

\bibitem{sqsup}
Castellanos-Beltran, M. A., Irwin, K. D., Hilton, G. C., Vale, L. R. \& Lehnert,  K. W. Amplification and squeezing of quantum noise with a tunable Josephson metamaterial, \textit{Nature Phys.} \textbf{4}, 929 - 931 (2008).


\bibitem{estsq1}
Milburn, G. J., Chen, W.-Y. \& Jones, K. R. Hyperbolic phase and squeeze-parameter estimation, \textit{Phys. Rev.
A} \textbf{50}, 801 (1994).

\bibitem{estsq2}
Chiribella, G., D'Ariano, G. M. \& Sacchi, M. F. Optimal estimation of squeezing, \textit{Phys. Rev. A} \textbf{73}, 062103 (2006).

\bibitem{estsq12}
Gaiba, R. \& Paris, M. G. Squeezed vacuum as a universal quantum probe, \textit{Phys. Lett. A} \textbf{373}, 934-939 (2009).

\bibitem{estsq7}
\v Safr\' anek, D., Lee, A. R. \& Fuentes, I. Quantum parameter estimation using multi-mode Gaussian states, 
\textit{New J. Phys.} \textbf{17}, 073016 (2015).

\bibitem{estsq10}
\v Safr\' anek, D. \& Fuentes, I. Optimal probe states for the estimation of Gaussian unitary channels, 
http://arxiv.org/abs/1603.05545 (2016).

\bibitem{estsq4}
Pinel, O., Jian, P., Treps, N., Fabre, C. \& Braun, D. Quantum parameter estimation using general single-mode Gaussian states, \textit{Phys. Rev. A} \textbf{88}, 040102(R) (2013).

\bibitem{estsq6}
Fiur\'a\v sek, J. Continuous-variable quantum process tomography with squeezed-state probes, \textit{Phys. Rev. A} \textbf{92}, 022101 (2015).

\bibitem{estsq13}
Sparaciari, C., Olivares, S. \& Paris, M. G. Gaussian state interferometry with passive and active elements, \textit{Phys. Rev. A} \textbf{93}, 023810 (2016).

\bibitem{estsq5}
Adesso, G. Gaussian interferometric power, \textit{Phys. Rev. A} \textbf{90}, 022321 (2014).

\bibitem{Rupp}
Ruppert, L., Usenko, V. C. \& Filip, R. Estimation of the covariance matrix of macroscopic quantum states, \textit{Phys. Rev. A} \textbf{93}, 052114 (2016).

\bibitem{Usen}
Usenko, V. C., Ruppert, L. \& Filip, R. Quantum communication with macroscopically bright nonclassical states, \textit{Opt. Exp.} \textbf{23}, 31534-31543 (2015).

\end{thebibliography}
\end{document}